\documentclass[manuscript]{emulateapj-rtx4}

\slugcomment{To appear in ApJ}

\shorttitle{Episodic accretion and low-mass star formation}
\shortauthors{Stamatellos, Whitworth \& Hubber }

\begin{document}

\title{The importance of episodic accretion for low-mass star formation}

\author{Dimitris Stamatellos, Anthony P. Whitworth}
\affil{School of Physics \& Astronomy, Cardiff University, 5 The Parade, Cardiff CF24 3AA, UK}
\and
\author{David A. Hubber}
\affil{Department of Physics \& Astronomy, University of Sheffield, Hounsfield Road, Sheffield S3 7RH, UK}
\affil{School of Physics and Astronomy, University of Leeds, Leeds LS2 9JT, UK}

\email{D.Stamatellos@astro.cf.ac.uk}

\begin{abstract}
A star acquires much of its mass by accreting material from a disc. Accretion is probably not continuous but  episodic. We have developed a method to include the effects of episodic accretion in simulations of star formation. Episodic accretion results in bursts of radiative feedback, during which a protostar is very luminous, and its surrounding disc is heated and stabilised. These bursts typically last only a few  hundred years. In contrast, the lulls between bursts may last a few thousand years; during these lulls the luminosity of the protostar is very low, and its disc cools and fragments. Thus, episodic accretion enables the formation of low-mass stars, brown dwarfs and planetary-mass objects by disc fragmentation. If episodic accretion is a common phenomenon among young protostars, then the frequency and duration of accretion bursts may be critical in determining the low-mass end of the stellar initial mass function.
\end{abstract}

\keywords{Stars: protostars, formation, low-mass, brown dwarfs -- Accretion, accretion disks --Methods: Numerical -- Hydrodynamics -- Radiative Transfer}

\section{Introduction}

Stars form from the collapse of gravitationally unstable cores in molecular clouds. Interstellar turbulence ensures that such cores have a net angular momentum, and so protostars form attended by accretion discs; much of the collapsing material first infalls onto the disc and then spirals into the protostar \citep{Terebey84}. If the disc cannot efficiently transport angular momentum outwards, accretion onto the central protostar is slow, and the disc grows in mass. The disc may then become gravitationally unstable and fragment \citep{Bate09b,Attwood09}, spawning secondary protostars, primarily of low-mass \citep{Stamatellos09,Stamatellos10}, i.e. low-mass hydrogen-burning stars ($0.08\,{\rm M}_\odot\;{\rm to}\,\sim0.2\,{\rm M}_\odot$), brown dwarfs ($0.012\,{\rm M}_\odot\;{\rm to}\;0.08\,{\rm M}_\odot$), and even planetary-mass objects ($\la 0.012\,{\rm M}_\odot$). This mechanism \citep{Whitworth06} is the only proposed mechanism that addresses in detail all the critical observational constraints, i.e. the shape of the low-mass end of the  initial mass function, the brown dwarf desert, the binary statistics of low-mass objects, and the formation of free-floating planetary-mass objects \citep{Stamatellos07b,Stamatellos09}. 

However, as mass flows towards the protostar, gravitational energy is transformed into thermal energy due to viscous dissipation in the disc and at the accretion shock around the protostar. It has been pointed out \citep{Krumholz06,Bate09,Offner09,Urban10,Krumholz10, Offner10}  that this accretion-related energy heats and stabilises the disc, thereby suppressing fragmentation and low-mass star formation.   More specifically, \cite{Bate09} finds that when one ignores radiative feedback too many brown dwarfs form in simulations of fragmenting clouds, whereas with radiative feedback a smaller number of brown dwarfs form. However, the simulations of \cite{Bate09} include radiative feedback due to the conversion of gravitational energy to thermal energy only down to the sink radius, i.e. 0.5~AU; thus, \cite{Bate09} simulations only account for a small fraction ($\sim 2\%$) of the total radiative feedback from young protostars and represent a lower limit on the effect of radiative feedback in disc fragmentation and low mass-star formation \citep[cf.][]{Offner09}. When the radiative feedback from young protostars is fully included then the formation of low-mass and brown dwarfs by disc fragmentation is almost fully suppressed \citep[][their Fig.~6]{Offner09}.  Furthermore, based on the effects of radiative feedback,  \cite{Offner10}  conclude that disc fragmentation is not a major contributor to low-mass stars and brown dwarfs. Moreover, disc fragmentation is currently the only model that predicts the statistics of low-mass stars and brown dwarfs \citep{Stamatellos09}. Hence, the formation of low-mass stars ($\stackrel{<}{_\sim}0.2$ M$_{\sun}$), which, if one includes brown dwarfs \citep{Whitworth07}, constitute $\,\sim\!60\%$ of all stars \citep{Kroupa01}, remains an open problem.

Previous studies have assumed that the accretion of material onto protostars, and hence their radiative feedback, is continuous \citep{Krumholz06,Bate09,Offner09,Krumholz10}. However, there is strong observational and theoretical evidence that accretion of material is often episodic \citep{Herbig77,Dopita78, Reipurth89, Hartmann96,Greene08,Peneva10}; it is concentrated in bursts that last from a few tens to a few hundred years. In particular, FU Ori-type stars exhibit sudden large ($\sim$5 mag) increases in their brightness, and then fade away over decades or centuries \citep{Hartmann96,Greene08,Peneva10}. The estimated accretion rates during these bursts are up to  $\sim 5\times 10^{-4}\,{\rm M}_{\sun}\,{\rm yr^{-1}}$, so each burst can deliver a few $10^{-2}\,{\rm M}_{\sun} $ onto the protostar.  Further evidence for episodic accretion comes from the periodically spaced knots seen in bipolar jets \citep[e.g.][]{Reipurth89}. Bipolar jets are driven off along the protostellar rotation axis by the energy released as material spirals in and accretes onto the protostar; periodically spaced knots therefore imply episodic accretion. Finally, the luminosity problem provides indirect observational support for episodic accretion. By the end of the Class 0 phase ($\sim10^5$ yr), half the protostar's final mass has been accumulated. Thus for a final mass of $1\,{\rm M}_\odot$, the mean accretion rate onto the protostar must be $\sim5\times 10^{-6}\,{\rm M}_{\sun}\,{\rm yr}^{-1}$, and the mean accretion luminosity must be $\sim 25\,{\rm L}_{\sun}$. This is much larger than the observed bolometric luminosities of typical solar-type protostars \citep{Kenyon90,Evans09}. Episodic accretion mitigates this problem, as the luminosity is only large intermittently, and most protostars are observed between bursts.

The cause of episodic accretion is uncertain \citep{Bonnell92,Bell94,Vorobyov05,Zhu07,Zhu09,Zhu09b,Zhu10, Zhu10b}, but the most compelling explanation \citep{Zhu09} involves the interplay between two different modes of angular momentum transport. The outer disc is sufficiently cool that it is susceptible to gravitational instability (GI); as well as creating low-mass companions in the disc, GI also results in gravitational torques that transport angular momentum outwards, thereby allowing material to spiral inwards towards the central protostar. In contrast, the inner disc  is too hot for GI, but, if it becomes hot enough, the gas is thermally ionised to a sufficient degree that it couples to the magnetic field and the magneto-rotational instability (MRI) kicks in; the MRI then transports angular momentum outwards, allowing matter to spiral in further and onto the central protostar. The upshot is that matter accumulates in the inner disc, and steadily heats up until it is sufficiently ionised to activate MRI, whereupon the accumulated matter spirals into the central protostar, giving rise to an outburst. Once most of the accumulated matter in the inner disc has been deposited onto the central protostar, the temperature of the remaining matter falls, the MRI shuts off, and the accumulation of matter resumes. 

In this paper, we present a semi-analytic prescription, which is based on the model of \cite{Zhu10}, to incorporate episodic accretion in radiation hydrodynamic simulations of star formation.  We perform simulations of the collapse of a star forming molecular cloud and show that episodic accretion plays a crucial role in promoting disc fragmentation and low-mass star formation.

\section{Computation Method}
\label{sec:methods}

We use the SPH code {\sc seren} \citep{Hubber11}, in which the chemical and radiative processes that regulate the gas temperature are treated with the flux-limited diffusion approximation of \cite{Stamatellos07} and \cite{Forgan09}. The net radiative heating rate for the particle $i$ is 
\begin{equation} 
\label{eq:radcool}
\left. \frac{du_i}{dt} \right|_{_{\rm RAD}} =
\frac{\, 4\,\sigma_{_{\rm SB}}\, (T_{_{\rm BGR}}^4-T_i^4)}{\bar{{\Sigma}}_i^2\,\bar{\kappa}_{_{\rm R}}(\rho_i,T_i)+{\kappa_{_{\rm P}}}^{-1}(\rho_i,T_i)}\,.
\end{equation}
The positive term on the right hand side represents heating by the background radiation field, and ensures that the gas and dust cannot cool radiatively below the background radiation temperature $T_{_{\rm BGR}}$. $\sigma_{_{\rm SB}}$ is the Stefan-Boltzmann constant, $\bar{{\Sigma}}_i$ is the mass-weighted mean column-density, and  $\bar{\kappa}_{_{\rm R}}(\rho_i,T_i)$ and ${\kappa_{_{\rm P}}}(\rho_i,T_i)$ are suitably adjusted Rosseland- and Planck-mean opacities. The method takes into account compressional heating, viscous heating, heating by the background radiation field, and radiative cooling/heating. The method has been extensively tested \citep{Stamatellos07,Stamatellos08}. In particular it reproduces the results of both detailed 3D simulations \citep{Masunaga00,Boss79,Whitehouse06}, and also -unlike other schemes -- analytic test calculations \citep{Spiegel57,Hubeny90}. The gas is assumed to be a mixture of hydrogen (70\%) and helium (30\%) , with  an equation of state \citep{Black75} that accounts (i) for the rotational and vibrational degrees of freedom of molecular hydrogen, and (ii) for the different chemical states of hydrogen and helium. For the dust and gas opacity we set  $\kappa(\rho,T)=\kappa_0\ \rho^a\ T^b\,$, where $\kappa_0$, $a$, $b$ are constants \citep{Bell94} that depend on the species and the physical processes contributing to the opacity at each $\rho$ and $T$, for example, ice mantle melting, the sublimation of dust, molecular and H$^-$ contributions.

Up to densities $\rho\sim 10^{-9}\,{\rm g}\,{\rm cm}^{-3}$, the self-gravitating gas dynamics, energy equation and associated radiation transport are treated explicitly. Wherever a gravitationally bound condensation with $\rho >10^{-9}\,{\rm g}\,{\rm cm}^{-3}$ is formed, it is presumed to be destined to form a protostar. Then, in order to avoid very small timesteps, it is replaced with a sink \citep{Bate95}, i.e.  a spherical region of radius $1\,{\rm AU}$. The sink  interacts with the rest of the computational domain only through its gravity and luminosity. Matter that subsequently flows into a sink, and is bound to it, is assimilated by the sink, and ultimately ends up in a protostar at the centre of the sink.  

The radiation from protostars that form during the simulation is taken into account by invoking a background radiation field with temperature $T_{_{\rm BGR}}({\bf r})$ (see Eq.~\ref{eq:radcool}) that is a function of position relative to all the protostars present in the simulation \citep{Stamatellos07,Stamatellos09b},
 \begin{eqnarray}
T_{_{\rm BGR}}^4({\bf r})&=&\left(10\,{\rm K}\right)^4+\sum_n\left\{\frac{L_n}{16\,\pi\,\sigma_{_{\rm SB}}\,|{\bf r}-{\bf r}_n|^2}\right\}\,,
\end{eqnarray}
where $L_n$ and  ${\bf r}_n$, are the luminosity and position of the $n${\footnotesize th} protostar.

At all times, the luminosity of a protostar  is given by
\begin{equation}
L_n=\left( \frac{M_n}{{\rm M}_{\sun}}\right)^3{\rm L}_{\sun}+\frac{f G M_n \dot{M}_n}{R_n}\,,
\end{equation}
where $M_n$ is the mass of the protostar, $R_n$ its radius, and $\dot{M}_n$ is the accretion rate onto it. The first term on the righthand side is the intrinsic luminosity of the protostar (due to contraction and nuclear reactions in its interior). The second term is the accretion luminosity. $f=0.75$ is the fraction of the accretion energy that is radiated away at the photosphere of the protostar, rather than being expended driving jets and/or winds \citep{Offner09}. We assume $R_n=3{\rm R}_{\sun}$ is the typical radius of a young protostar \citep{Palla93}. At the initial stages of star formation the accretion luminosity dominates over the intrinsic  luminosity of the protostar.

\section{A phenomenological time-dependent model of episodic accretion}
\label{sec:ea.model}

Simulations of self-gravitating hydrodynamics on the scale of molecular cloud cores (i.e. sizes from $10^4$ to $10^5\,{\rm AU}$ and masses from $0.3$ to $10\,{\rm M}_\odot$) can achieve sufficient resolution to capture the formation of discs around young protostars, and the effects of GIs in the outer regions of such discs. However, such simulations cannot capture what happens in the inner disc region, where sinks are invoked.  It is normally assumed that any matter inside a sink flows instantaneously onto the central protostar. Here, we assume that  the mass accreted into a sink  is deposited on an inner accretion disc (IAD) inside the sink, where it piles up until it becomes hot enough that thermal ionisation couples the matter to the magnetic field. At this juncture the MRI is activated, transporting angular momentum outwards, and thereby allowing the matter accumulated in the IAD to spiral inwards and onto the central protostar. Hence the sink mass is divided between the central protostar ($M_\star$) and its IAD ($M_{_{\rm IAD}}$):
\begin{equation}
M_{_{\rm SINK}}=M_{\star}+M_{_{\rm IAD}}\,.
\end{equation}
Matter is assumed to flow from the IAD onto the protostar at a rate
\begin{equation}
\dot{M}_{\star}=10^{-7}\,{\rm M}_{\sun}\,{\rm yr}^{-1}+\dot{M}_{_{\rm MRI}}\,,
\end{equation}
where the first term on the righthand side is a low regular accetion rate that obtains at all times, and the second term is a much higher accretion rate that obtains only when MRI acts to transport angular momentum in the IAD.  The IAD is presumed to be a continuation of the much more extended accretion disc outside the sink, which is simulated explicitly.

In detailed disc models \citep{Zhu09,Zhu09b,Zhu10,Zhu10b} the matter in the IAD couples to the magnetic field, due to thermal ionisation, once the temperature reaches a threshold of $T_{_{\rm MRI}}\sim 1400\,{\rm K}$.  Using an $\alpha$-parameterization \citep{Shakura73} for the effective viscosity delivered by the MRI, \cite{Zhu10} then estimate that the accretion rate during an outburst is
\begin{equation}
\label{eq:mrimdot}
\dot{M}_{_{\rm MRI}}\sim 5\times10^{-4}\,{\rm M}_{\sun}\,{\rm yr}^{-1}\,\left(\frac{\alpha_{_{\rm MRI}}}{0.1}\right)\,,
\end{equation}
where $\alpha_{_{\rm MRI}}$ (where $\alpha_{_{\rm MRI}}$ is the effective Shakura-Sunyayev  parameter \citep{Shakura73} for MRI viscosity).  \cite{Zhu10} also find that the duration of an outburst is
\begin{eqnarray}
\label{eq:mridt}
\Delta t_{_{\rm MRI}} &\,\sim\,0.25\,&{\rm kyr}\,\left(\frac{\alpha_{_{\rm MRI}}}{0.1}\right)^{-1}\,
\left(\frac{M_{\star}}{0.2{\rm M}_{\sun}}\right)^{2/3}\times\nonumber\\ 
& &
\times\left(\frac{\dot{M}_{_{\rm IAD}}}{10^{-5}\ {\rm M}_{\sun}\,{\rm yr}^{-1}}\right)^{1/9}\,.
\end{eqnarray}
Here $\dot{M}_{_{\rm IAD}}$ is the rate at which matter flows into the sink and onto the IAD.
We assume that the critical temperature  for the MRI to be activated is reached when enough mass for an MRI-enabled outburst has been accumulated in the IAD, i.e. 
\begin{equation}
M_{_{\rm IAD}}\;\,>\;\
M_{_{\rm MRI}}
\;\,\sim\;\,{\dot{M}_{_{\rm MRI}}}{\Delta t_{_{\rm MRI}}}\,.
\end{equation}
Using Eqs.~\ref{eq:mrimdot} and \ref{eq:mridt} we obtain
\begin{equation}
M_{_{\rm IAD}}\!>\! 0.13\,{\rm M}_{\sun}\!
\left(\frac{M_{\star}}{0.2{\rm M}_{\sun}}\right)^{2/3}\!
\left(\frac{\dot{M}_{_{\rm IAD}}}{10^{-5}\,{\rm M}_{\sun}\,{\rm yr}^{-1}}\right)^{1/9}.
\end{equation}

We do not model in detail the thermal and ionisation balance in the IAD, since this is not the main concern of this paper. Instead, based on observations and models of FU Ori-type stars \citep{Hartmann96,Zhu10}, we presume that rapid accretion onto the protostar, at a rate given by
\begin{eqnarray}\label{w}
\dot{M}_{\rm MRI}\!=\!\frac{M_{_{\rm MRI}}}{\Delta t_{_{\rm MRI}}}\!\exp\!\left\{\!-\!\frac{(t\!-\!t_0)}{\Delta t_{_{\rm MRI}}}\right\},t_{_{\rm O}}\!<\!t\!<\!t_{_{\rm O}}\!+\!\Delta t_{_{\rm MRI}}
\end{eqnarray}
is initiated as soon as $M_{_{\rm IAD}}$ exceeds $M_{_{\rm MRI}}$; $t_{_{\rm O}}$ is the time at which this occurs, and the outburst terminates at $t_{_{\rm O}}+\Delta t_{_{\rm MRI}}$. 

The time taken to accumulate (or re-accumulate) $M_{_{\rm MRI}}$ is $\Delta t_{_{\rm ACC}} \sim M_{_{\rm MRI}}/\dot{M}_{_{\rm IAD}}$, i.e. 
\begin{equation}\label{EQN:dtACC}
\Delta t_{_{\rm ACC}}\!\simeq\!13{\rm kyr}\!\left(\!\frac{M_\star}{0.2{\rm M}_\odot}\!\right)^{2/3}\!\left(\!\frac{\dot{M}_{_{\rm IAD}}}{10^{-5}{\rm M}_\odot\,{\rm yr}^{-1}}\!\right)^{-8/9}\,.
\end{equation}
This is much longer than the duration of the outburst (see Eq.~\ref{eq:mridt}).

With the above formulation the only free parameter is $\alpha_{_{\rm MRI}}$, which controls the strength and duration of the outburst; increasing $\alpha_{_{\rm MRI}}$ makes for a more intense, shorter outburst. The mass delivered in each outburst, and the duration of the interval between outbursts, are essentially independent of  $\alpha_{_{\rm MRI}}$, which is fortunate, in the sense that $\alpha_{_{\rm MRI}}$ is rather uncertain. Observations and  simulations suggest  $\alpha_{_{\rm MRI}}\!=\!0.01\;{\rm to}\;0.4$ \citep{King07}. We have adopted $\alpha_{_{\rm MRI}}\!=\!0.1\,$, which is consistent with observations of FU Ori events \citep{Zhu07}.

\begin{figure*}[t]
\centerline{
\includegraphics[height=17.5cm,angle=-90]{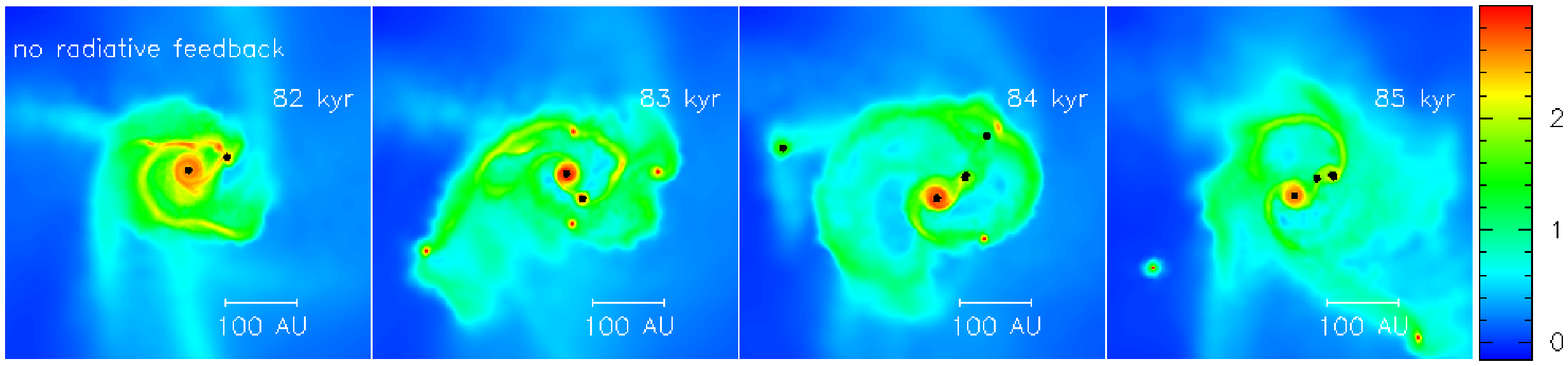}}
\caption{Evolution of the accretion disc around the primary protostar forming in a collapsing turbulent molecular cloud core, without radiative feedback from the protostar. The disc around the primary protostar increases in mass, becomes gravitationally unstable, and fragments to form 3 low-mass stars, 2 brown dwarfs and 2 planetary-mass objects.  The colour encodes the logarithm of column density, in ${\rm g\ cm}^{-2}$. }
\label{fig1}
\centerline{
\includegraphics[height=17.5cm,angle=-90]{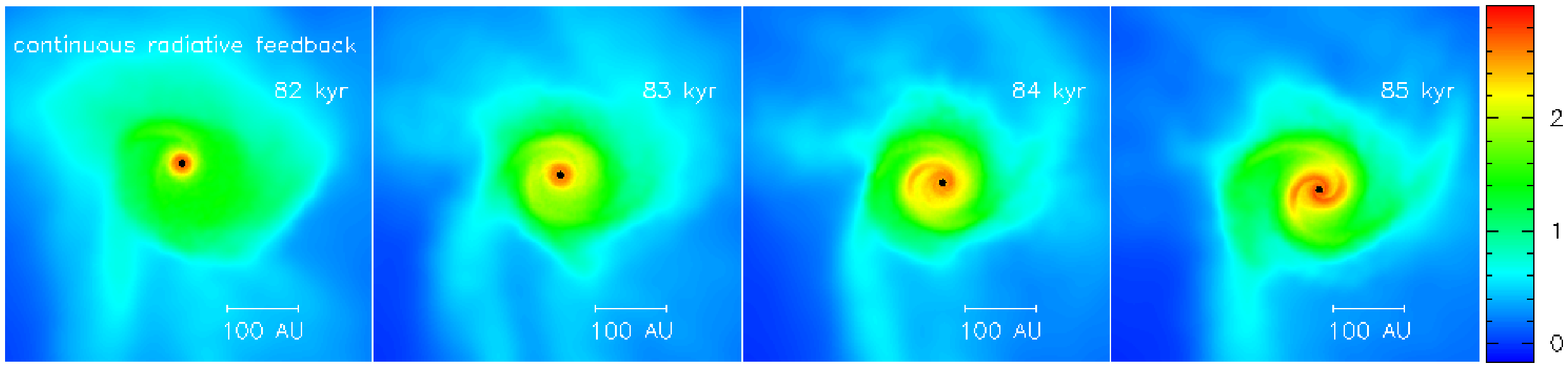}}
\caption{Evolution of the accretion disc around the primary protostar with  continuous accretion, continuous radiative feedback from the protostar. The disc grows in mass, but radiative feedback makes it so hot that it does not fragment.}
\label{fig2}
\centerline{
\includegraphics[height=17.5cm,angle=-90]{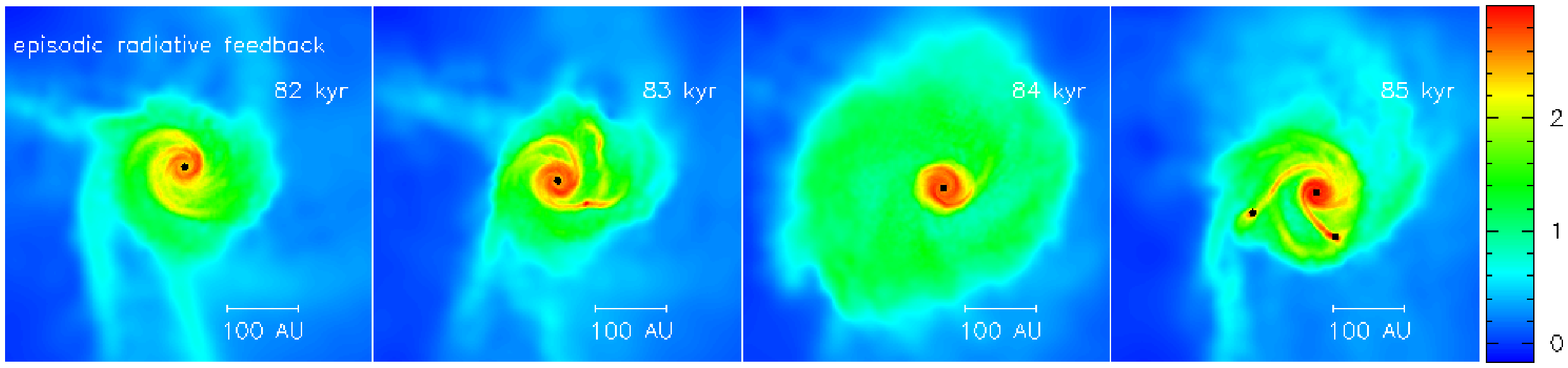}}
\caption{Evolution of the accretion disc around the primary protostar with episodic accretion and episodic radiative feedback from the protostar. The disc becomes gravitationally unstable (first and second column) but fragmentation is damped by heating due to an accretion burst (third column); however, after this burst the disc cools sufficiently to undergo gravitational fragmentation; 2 low-mass stars form in the disc.}
\label{fig3}
\end{figure*}

\section{The importance of episodic accretion in low-mass star formation}
\label{sec:results}

To evaluate the consequences of episodic accretion for disc fragmentation and low-mass star formation , we perform radiation hydrodynamic simulations of a collapsing turbulent molecular core, with properties chosen to match  the observed properties of prestellar cores \citep[e.g.][]{Andre00}. The initial density profile is
\begin{eqnarray}
\rho(r)&=&\frac{\rho_{_{\rm KERNEL}}}{\left[1\,+\,\left(r/R_{_{\rm KERNEL}}\right)^2\right]^2}\,.
\end{eqnarray}
Here $\rho_{_{\rm KERNEL}}=3\times 10^{-18}\,{\rm g}\,{\rm cm}^{-3}$ is the central density, and $R_{_{\rm KERNEL}}=5,000\,{\rm AU}$ is the radius of the central region within which the density is approximately uniform. The outer envelope of the core extends to $R_{_{\rm CORE}}=50,000\,{\rm AU}$, so the total core mass is $M_{_{\rm CORE}}=5.4\,{\rm M}_\odot$. The gas is initially isothermal at $T=10\,{\rm K}$, and hence the initial ratio of thermal to gravitational energy is $\alpha_{_{\rm THERM}}=0.3\,.$ We impose an initial random, divergence-free, turbulent velocity field, with power spectrum $P_kdk\propto k^{-4}dk$, to match the scaling laws observed in molecular clouds \citep{Larson81}, and amplitude such that $\alpha_{_{\rm TURB}}\equiv{U_{_{\rm TURB}}}/{|U_{_{\rm GRAV}}|}=0.3\,$.

 The core is assumed to evolve in isolation, i.e. any interactions with its environment are ignored. This appears to be the case  in Ophiuchus, where \cite{Andre07} conclude that individual cores  do not have time to interact with one another before evolving into protostars. In denser star formation regions interactions between cores may occur, but only very close encounters will influence disc fragmentation in scales less than a few hundred AU. Such encounters are expected to enhance disc fragmentation \citep{Thies10}.

The molecular cloud core is represented by $10^{6}$ SPH particles, so each SPH particle has mass $m_{_{\rm SPH}}\simeq 5\times 10^{-6}\,{\rm M}_\odot$. The minimum resolvable mass is therefore $M_{_{\rm MIN}}\simeq {\cal N}_{_{\rm NEIB}}m_{_{\rm SPH}}\simeq 3\times 10^{-4}\,{\rm M}_\odot$. This is much smaller than the thoretical minimum mass for star formation ($\sim 3\times 10^{-3}\,{\rm M}_\odot$; usually refered to as {\it The Opacity Limit}), and therefore all self-gravitating condensations should be reasonably well resolved.

\begin{figure*}[t]
\centerline{
\includegraphics[width=12cm,angle=-90]{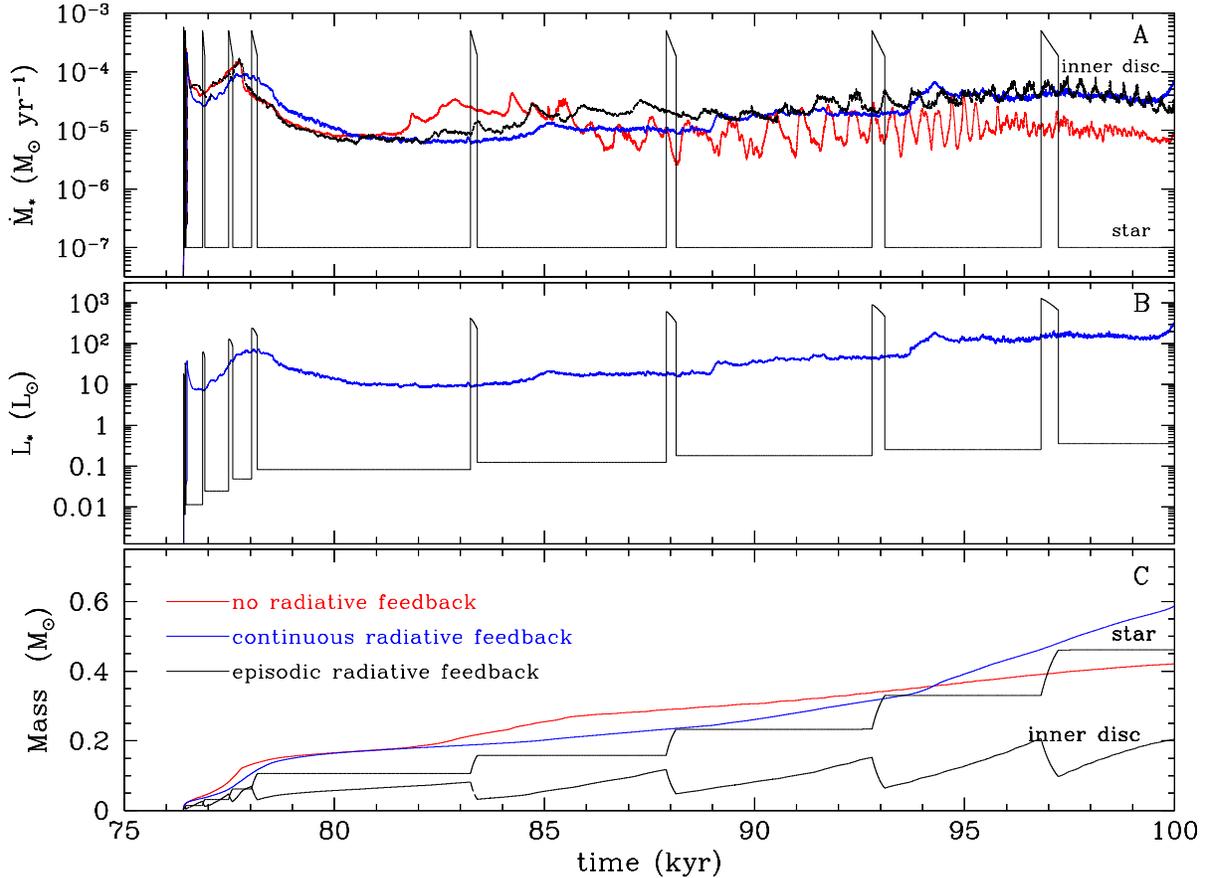}}
\caption{The evolution of the primary protostar and its IAD. Red lines are used for the case with accretion but no radiative feedback, blue lines for the case with continuous accretion and feedback, and black lines for the case with episodic accretion and feedback. (A) shows the accretion rate into the sink and the accretion rate onto the primary protostar. For the first two cases these two quantities are the same (by definition), but for episodic accretion they are different. (B) shows the accretion luminosities for the cases with continuous (blue) and episodic (black) accretion; the luminosity is effectively zero for the case with no radiative feedback. (C) shows the mass of the primary protostar, which increases very steadily for the runs with no radiative feedback (red), and with continuous radiative feedback (blue). For the case with episodic radiative feedback (black), we plot the mass of the protostar  and the mass of the IAD; the mass of the protostar only increases significantly during the brief accretion bursts.}
\label{fig4}
\end{figure*}

We perform three simulations, all with the same initial conditions, and differing only in their treatments of the luminosities of protostars. In all three simulations the initial collapse leads to the formation of a primary protostar (i.e. a first sink) at $\,t\!\sim\!77\,{\rm kyr}$, and this quickly acquires an extended accretion disc. The simulations only diverge after this juncture.

\subsection{No radiative feedback}

In the first simulation there is no radiative feedback from the  protostar, and the gas in the disc is only heated by compression and by viscous dissipation in shocks. It is therefore cool enough to experience strong GI \citep{Stamatellos09}. The resulting gravitational torques transport angular momentum outwards, allowing matter to spiral inwards and onto the primary protostar at a rate $10^{-5}\;{\rm to}\;10^{-4}\,{\rm M}_\odot\,{\rm yr}^{-1}$, but in the outer parts of the disc, at radii $R\ga 50\,{\rm AU}$, the GI also becomes highly non-linear locally, resulting in the formation of seven secondary protostars, with masses ranging from $0.008\,{\rm M}_\odot$ to $0.24\,{\rm M}_\odot$. This is illustrated in the sequence of frames on Fig.~1. The accretion rate onto the primary protostar and the growth of its mass are shown on Fig.~4 (red lines).

\subsection{Continuous radiative feedback}

In the second simulation, we assume that the matter entering a sink is immediately accreted onto the protostar at its centre. This results in an  accretion luminosity which is typically $10\;{\rm to}\;100\,{\rm L}_\odot$. It therefore heats the surrounding disc, and as a result GI saturates, generating low-amplitude spiral waves that  transport angular momentum outwards and allow matter to spiral into the protostar, but entirely suppressing disc fragmentation, at all radii. This is illustrated in the sequence of frames on Fig.~2. At all times the spiral structure in the disc is diffuse and has low amplitude, there is no fragmentation and therefore no secondary protostars form. The accretion rate onto the primary protostar, its resulting accretion luminosity, and the  growth of its mass are shown on Fig.~4 (blue lines).

\subsection{Episodic radiative feedback}

In the third simulation, we use the semi-analytic, time-dependent model for  angular momentum transport inside the sink that is described in Sect.~\ref{sec:ea.model}. The luminosity of the protostar during the MRI-driven accretion bursts is very high, and stabilises the disc against fragmentation. However, as the mass of the protostar ($M_\star$) builds up, the mass in the IAD that is required to activate MRI ($M_{_{\rm MRI}}$) also goes up, and at the same time the accretion rate into the sink ($\dot{M}_{_{\rm IAD}}$) goes down, so that the intervals between bursts become longer. During these intervals the luminosity of the protostar is relatively low, hence the disc cools down. Within a few kyr of the formation of the primary protostar, the interval between successive accretion bursts has increased to $\sim 5\,{\rm kyr}$, and this is sufficient time to allow the outer disc at $R\sim 50\;{\rm to}\;150\,{\rm AU}$, to undergo gravitational fragmentation. This is illustrated on Fig.~3. In the first two frames, at $82$ and $83\,{\rm kyr}$, the disc is quite unstable and tries to fragment, but then at $84\,{\rm kyr}$ it is heated by an accretion outburst and stabilised. Once the outburst is over, the disc cools back down, and fragments to produce two low-mass hydrogen-burning secondaries. (Other, statistically similar simulations also produce brown dwarfs and even planetary-mass objects, as in \citealt{Stamatellos09}). The accretion rates onto the primary protostar and onto its IAD, its resulting accretion luminosity, and the growth of the mass of the protostar and its IAD are shown on Fig.~4 (black lines).

\section{Discussion}
\label{sec:discussion}
These results demonstrate the importance of episodic accretion in facilitating the formation of low-mass secondary stars and brown dwarfs by disc fragmentation. The critical factor that determines whether disc fragmentation occurs is the time interval between successive accretion bursts. In the phenomenological model presented here, which is based on the detailed model of \cite{Zhu10}, this interval is given by Eq.~ \ref{EQN:dtACC}, and depends on the rate of inflow into the IAD, $\dot{M}_{_{\rm IAD}}$, and the mass of the primary protostar, $M_\star$. It is initially quite short, because $M_\star$ is small and $\dot{M}_{_{\rm IAD}}$ is large. However, over time $M_\star$ increases, and in general $\dot{M}_{_{\rm IAD}}$ decreases. By the time $M_\star\sim 0.2\,{\rm M}_\odot$ and $\dot{M}_{_{\rm IAD}}\sim 10^{-5}\,{\rm M}_\odot\,{\rm yr}^{-1}$, the interval is $\,\sim\!13\,{\rm kyr}$. This is sufficient time for the outer disc to fragment, since the condensation timescale for an unstable fragment at $50$ to $150\,{\rm AU}$ is $\la 5\,{\rm kyr}$. It follows that the relative abundance of very low-mass stars may be different in different star forming environments. In dense molecular cloud cores  where the rate of accretion onto the IAD is expected to be high, due to the large reservoirs of gas available, the interval between accretion bursts may tend to be shorter, and hence the opportunities for disc fragmentation less frequent. This should be reflected in a paucity of very low-mass stars  and brown dwarfs in such regions, relative to more diffuse star forming regions. 

We conclude that if episodic accretion is a common phenomenon among young protostars as observational and theoretical evidence suggests \citep{Herbig77,Dopita78, Reipurth89, Hartmann96,Greene08,Peneva10}, it may limit the effect of the luminosity of a protostar on its environment and create favourable conditions for disc fragmentation to occur. As disc fragmentation is predominantly a mechanism for forming low-mass stars and brown-dwarfs, episodic accretion may then have a significant influence on the lower end of the initial mass function.

\acknowledgments

We  thank S.~Goodwin, S.~Walch, and R.~W\"unsch for  reviewing the manuscript.  Simulations were performed using the Cardiff HPC Cluster {\sc merlin}.  Column density plots were produce using {\sc splash} \citep{Price07}. D.S. and A.W. acknowledge support by the STFC grant PP/E000967/1. DAH receives support from STFC and a Leverhume Trust Research Project Grant (F/00 118/BJ).

\bibliography{bibliography}{}

\end{document}